\documentclass[10pt, conference, letterpaper]{IEEEtran}
\ifCLASSINFOpdf
\else
\fi
\usepackage{xcolor}
\usepackage{comment}
\usepackage{graphicx}
\usepackage{subfigure}
\usepackage{booktabs}
\usepackage{mathtools}

\usepackage{color, colortbl}
\definecolor{dark green}{RGB}{94,145,40}

\newcommand{\ennote}[1]{\textit{\textcolor{orange}{#1 --EN}}}
\providecommand{\ie}{\emph{i.e.,} }
\providecommand{\eg}{\emph{e.g.,} }

\newcommand\mypara[1]{\noindent \textbf{#1}}

\providecommand{\ione}{\emph{(i)} }
\providecommand{\itwo}{\emph{(ii)} }
\providecommand{\ithree}{\emph{(iii)} }
\providecommand{\ifour}{\emph{(iv)} }
\providecommand{\ifive}{\emph{(v)} }

\usepackage{array}

\usepackage{biblatex}
\usepackage{tikz}
\newcommand\ChapterPrecis[2]{%
\begin{tikzpicture}[remember picture,overlay]
\node[anchor=north, draw=black, fill=yellow!20, rounded corners, yshift=-#1] at (current page.north)
{\parbox[t][1.5cm][c]{\textwidth}{#2}};
\end{tikzpicture}%
}

\begin{document}

%
\title{A Game of NFTs: Characterizing NFT Wash Trading in the Ethereum Blockchain}





\author{\IEEEauthorblockN{Massimo La Morgia,
Alessandro Mei,
Alberto Maria Mongardini, and
Eugenio Nerio Nemmi,~\IEEEmembership{Fellow,~IEEE}}
\IEEEauthorblockA{Department of Computer Science, University of Sapienza, Rome}
\IEEEauthorblockA{Email: \{lamorgia, mei, mongardini, nemmi\}@di.uniroma1.it}
}

\markboth{Journal of \LaTeX\ Class Files,~Vol.~14, No.~8, August~2015}%
{Shell \MakeLowercase{\textit{et al.}}: Fake, Clone channels and other Fantastic Beasts}
\IEEEtitleabstractindextext{%
\begin{abstract}
The Non-Fungible Token (NFT) market in the Ethereum blockchain experienced explosive growth in 2021, with a monthly trade volume reaching \$6 billion in January 2022. However, concerns have emerged about possible wash trading, a form of market manipulation in which one party repeatedly trades an NFT to inflate its volume artificially.
Our research examines the effects of wash trading on the NFT market in Ethereum from the beginning until January 2022, using multiple approaches. We find that wash trading affects 5.66\% of all NFT collections, with a total artificial volume of \$3,406,110,774.
We look at two ways to profit from wash trading: Artificially increasing the price of the NFT and taking advantage of the token reward systems provided by some marketplaces.
Our findings show that exploiting the token reward systems of NFTMs is much more profitable (mean gain of successful operations is \$1.055M on LooksRare), more likely to succeed (more than 80\% of operations), and less risky than reselling an NFT at a higher price using wash trading (50\% of activities result in a loss). Our research highlights that wash trading is frequent in Ethereum and that NFTMs should implement protective mechanisms to stop such illicit behavior.
\end{abstract}
}

\maketitle

\ChapterPrecis{0.1cm}{If you cite this paper, please use the ICDCS reference:
M. La Morgia, A. Mei, A. M. Mongardini, and E. N. Nemmi. "A Game of NFTs: Characterizing Nft Wash Trading in the Ethereum Blockchain." \textit{2023 IEEE 43rd International Conference on Distributed Computing Systems (ICDCS). IEEE, 2023}.}

\IEEEdisplaynontitleabstractindextext

%
\IEEEpeerreviewmaketitle

\section{Introduction}
%
%
%
%
\IEEEPARstart{}{} 
A non-Fungible Token (NFT) is a blockchain token designed to represent the ownership of a unique digital or physical asset. The scene of digital art and the NFTs on Ethereum saw impressive growth in recent years, with an increment of the trading volume from \$33 million in 2020 to \$13 billion in 2021~\cite{nft_trading_volume_2021}. 
To understand the hype behind this technology, Twitter CEO Jack Dorsey sold the screenshot of its first Twitter Tweet as an NFT for more than 2.9 million USD~\cite{first_tweet}.
In response to this broad interest, several NFT marketplaces (NFTMs) emerged to ease the complex interaction with the blockchain and resolve the trust issue between the buyer and the seller.

Among the NFTMs, the most popular one is OpenSea, with a trading volume of more than 12.5 billion in 2021~\cite{nft_trading_volume_2021}.

However, where money flows, malicious actors enter the scene and try to exploit the technology to their advantage. 
Indeed, due to the unregulated nature of blockchains, market manipulations, illegal on the traditional stock market, are common in the crypto market.
One of these manipulations is wash trading, a fraud in which different colluding actors artificially increase the volume of an asset.
In this work, we focus on wash trading manipulations involving ERC-721 compliant NFTs on the Ethereum blockchain, characterizing 
the phenomenon both economically and temporally. 
To this end, we parse the Ethereum blockchain collecting all ERC-721 NFT transfers.
We use different techniques to confirm potential wash trading events finding that this manipulation of the market is responsible for a volume of 3.4 billion US dollars. 
We also discover that wash trading activities are especially prevalent on marketplaces that offer token reward systems based on the user's trading volume.
Among the highly affected marketplaces by wash trading activities, we find that more than 84\% of the volume of the marketplace LooksRare is due to wash trading.
%
Wash trading operations are typically performed in a short time window, with 25.98\% of them lasting only one day and 51.67\% less than ten days. 

Lastly, we analyze if the wash trading activity is profitable. In particular, we divided the activities into two sets. The first is the set of wash trading activities that aims to profit by exploiting the reward systems of the marketplaces. The second is the set of activities that, through wash trading, attempt to increase the perceived value of a particular NFT and resell it at a higher price.
%
%
Our data suggest that it is much more profitable to exploit the token reward system rather than try to increase the value of an NFT.
%

Our key contributions are:
\begin{itemize}
    \item \textbf{Systematic analysis of wash trading on Ethereum.}
    We gather all ERC-721 NFTs of Ethereum from its inception to Jan. 18, 2022, collecting more than 34M assets whose trades moved a volume of more than 34B US dollars.
    We leverage our dataset to find wash trading activities, discovering that approximately 10\% of the global NFTs trading volume has been generated through wash trading manipulations. 
    \item \textbf{Comparison of wash trading detection methodologies.}
    We compare different methodologies to confirm wash trading activities.
    Combining these methodologies, we uncover how wash traders operate. In particular, we find that most of the colluding traders receive funds from a common account and then send the profit to another common account.
    %
    \item \textbf{Wash Trading characterization.}
    We show that most wash trading events last less than ten days, usually occur near the launch of a new collection, and are performed by only two accounts doing ``round-trip'' trading of the NFT. We discover the presence of "serial wash traders", as 27.16\% of the accounts involved in NFT wash trading participate in 72.93\% of activities.
    Moreover, we quantify the amount of artificial volume on the six major NFT marketplaces in Ethereum.
    
    \item \textbf{Analysis on the profitability of wash trading.} To the best of our knowledge, we are the first to measure if the wash trading activity on NFTs can be profitable. We analyze gains and losses of these events either when entities exploit the token rewards system of the NFTMs or when they resell the asset after the wash trading operations. 
    We find that while reselling the NFTs is risky (50\% of the event ended with a loss), exploiting the reward systems is much more profitable, with almost 80\% of the events resulting in a gain.
\end{itemize}

\section{Background}
\subsection{The Ethereum blockchain}
Ethereum is a proof-of-work blockchain that uses the Ether (ETH) as its native coin~\cite{buterin2014next}. It is the second-largest blockchain per market cap, with more than 180 billion US dollars~\cite{marketcap}, just below Bitcoin. One of the reasons for its success is the introduction of smart contracts, pieces of code that are decentralized and executed directly on-chain. Smart contracts 
make decentralized applications (dApp) possible and are fundamental to the so-called Web 3.0.
Through a smart contract, it is possible to create new assets on Ethereum, such as tokens or Non-Fungible Tokens (NFTs).

Two types of accounts can be used to interact with the Ethereum blockchain: Externally Owned Accounts (EOAs) and contract accounts. 
Both kinds of accounts are represented by a public address of 42 characters, representing the Keccak-256 hash~\cite{dworkin2015sha} of its public key.
The EOAs are managed by individuals, while the latter by smart contracts. Altought both types of accounts can hold tokens and NFTs, the smart contract account can manage its assets only in response to a transaction. 
Indeed, transactions are how accounts can interact with Ethereum and are used to change its state.
In Ethereum, a transaction can be used to transfer ETH and assets, such as tokens or NFTs, or to invoke a smart contract. However, making a transaction has a cost, the \textit{gas fee}, that is measured in Gwei ($1 \times 10^{-9}$ ETH). This fee can change dramatically based on the network's congestion and the complexity of the transaction.
Moreover, transactions can interact with a smart contract calling one of its functions that triggers the smart contract logic.

\subsection{NFT}
One particular type of asset in Ethereum is the Non-Fungible Token (NFT). 
Contrary to ERC-20 tokens, in which all tokens have the same properties and two tokens are indistinguishable from each other, each NFT is unique, having specific properties that distinguish it from the others.
In Ethereum, the most used standard to represent NFTs is ERC-721~\cite{erc-721}. This standard defines the structure of the smart contract that generates the NFT assets, specifying a set of functions and events that must be implemented. 
Each smart contract usually manages a \textit{collection} of NFTs, namely the set of NFTs created (minted) by the same smart contract. In particular, each NFT has a token ID identifying it within its collection. We can uniquely identify an NFT by using its smart contract address and its token ID as a tuple.

\subsection{NFTM}
The simplest way to buy an NFT is to use an NFT Marketplace (NFTM). NFTMs are platforms where users can buy and sell NFTs of different collections, interacting with the smart contract of the marketplace. Some marketplaces need the NFT on sale to be kept in a particular account called \textit{account escrow}, managed by the marketplace. The account escrow holds the asset on behalf of the seller until it is sold.
To offer their services, marketplaces collect fees for each transaction, which usually is between 1 and 2.5\% for both buyer and seller. 
To incentivize the use of their platform, some marketplaces implement a reward system that compensates users based on the volume of their transactions.
We can analyze transactions made on the NFTMs by searching for their smart contract address as the contract that transactions interact with.

\subsection{Wash trading}
\label{sec:wash_trading}
Wash trading is a form of market manipulation in which a set of (colluding) users trade the same asset to create artificial activity.
On the stock market, wash trading is usually performed by simultaneously placing a sell and buy order on the same asset without incurring market risk or changing the trader's market position; it is also referred to as \textit{Round Trip Trading}.
The purpose of wash trading is to influence a specific asset's pricing or trading activities, generating interest in the asset and attracting external traders.
This practice has been illegal for almost a century in the U.S. stock markets since the federal Commodity Exchange Act in 1936~\cite{commodityact}.
However, in the unregulated crypto market, this is not true, and in the NFT community, people believe wash trading activity significantly impacts NFTs' trading volume~\cite{significant_wash}.
In the NFT ecosystem, the main concern is that wash trading can create the illusion that there is market interest in the target NFTs. The goal is to increase their value artificially.
Furthermore, since some NFTMs reward users according to their trading volumes, wash trading can also be used to claim more undeserved reward tokens.

\section{The dataset}
This section describes the approach used to build the dataset and its main properties. 
\subsection{Building the dataset}
\label{sec:bulding_dataset}

To conduct our study, we analyze the Ethereum blockchain by collecting the most extensive list of NFTs that might have been the object of wash trading activities. In particular, we parse the Ethereum blockchain from its first block (mined on Jul 30, 2015) to block \# 14,030,326 (mined on Jan 18, 2022). Due to the size of the Ethereum blockchain, the best strategy is to run an Ethereum full node locally in our laboratory so that queries execute quickly. We used Geth~\cite{go_ethereum}, the official Go implementation of the Ethereum protocol, and the Web3 Python~\cite{web3_py} library to query the blockchain. Specifically, Web3 is a library that allows interaction with a local or remote Ethereum node using HTTP, IPC, or WebSocket. Below we detail the approach used to build the dataset.

\mypara{Transfer events collection.} 
According to the ERC-721 standard, a smart contract that manages NFTs has to implement the functionality to transfer an NFT from one account to another through a method called Transfer. The method takes three parameters as input: The address that owns the NFT, the address of the recipient, and the ID of the NFT. Moreover, the standard requires that this Transfer method must emit an Event containing 
some parameters every time the smart contract performs a transfer (an Event is a signal mechanism used in Ethereum to track the internal state changes of the smart contracts). In the following, we refer to this kind of Event as a Transfer event. In particular, the ERC-721 standard requires that each transfer of this type emits an event that logs four arguments: the Keccak of the function name -- transfer --, the source address, the recipient address, and the token ID of the NFT. This event allows us to distinguish ERC-721 token from other kinds of tokens as ERC-20 and ERC-1155 (another standard that can be used to implement both fungible and not fungible tokens). 
Indeed the ERC-20 event emits a log of three arguments, while the ERC-1155 utilizes a function with a different Keccak.
In particular, the signature of the ERC-721 transfer events is \texttt{ddf252ad}.
By using this approach, we searched for Transfer events compliant with the ERC-721 standard and collected 52,871,559 Transfer events executed by 26,737 different smart contracts.
    
    
    
    \mypara{ERC-721 compliance.} 
    Emitting an ERC-721 compliant Transfer event does not guarantee that the smart contract complies with the standard. 
    Indeed, if the smart contract implements other methods, these may not comply with the ERC-721 specifications.
    To verify the compliance, we perform an additional step.
    However, determining the type of token implemented by a smart contract is still an open challenge~\cites{chen2019tokenscope, di2021identification}. Although several works dealt with this problem from the perspective of ERC-20 smart contracts, there are no proposed solutions for ERC-721. 
    Inspired by the works of~\cites{chen2020traveling, victor2019measuring}, we leverage the fact that an ERC-721 smart contract must extend the ERC-165 standard, which requires implementing the \textit{supportsInterface} method. This method takes an interface identifier (\eg \texttt{0x01ffc9a7}) as input and specifies whether or not the smart contract implements the target interface. Thus, we use the supportsInterface method specifying in input the ERC-721 interface-id (\texttt{0x80ac58cd}).
    Using this technique, we discover that 25,878 out of the 26,737 (96.8\%) retrieved smart contracts are ERC-721 compliant.
    


At the end of the process, for each NFT collected, we stored the NFT smart contract address, the NFT tokenID, and the list of all the transfer events involving the NFT. Then, for each transfer event, we store the source, the recipient, and the transaction hash that we use to retrieve all the transaction-related information, such as the block number, the gas fee, and the number of tokens transferred.

\mypara{Collecting standard transactions.}
As the last step, we query our node a second time to retrieve all the transactions (sent and received) for accounts that appear as the source or the recipient of a Transfer event.
%
For each collected transaction, we stored the transaction hash, the block number, the recipient, the amount of ERC-20 token or Ether transferred, and, in the case of an ERC-20 transfer, its smart contract.  




\subsection{Dataset overview}
To better contextualize the wash trading activities, we first explore our dataset and provide an overview of the NFTs ecosystem on the Ethereum blockchain.

We collected 25,878 smart contracts that transferred 34,754,447 different NFTs and moved over 388 billion USD. However, part of this activity is not related to art, collectibles, or, in any case, NFTs that may be subject to wash trading. For example, UniswapV3, a Decentralized Exchange (DEX) in Ethereum, generates an NFT every time a user deposits an asset (usually crypto). The NFT represents the position and can be redeemed by the owner to get tokens back. UniswapV3 accounts for some 0.2M NFTs and 91\% of the volume. Clearly, these NFTs are not interesting for our study, so even though they are part of the dataset, they have no impact on the rest of this work.

Some of the NFTs are traded in NFT markets. To analyze the activities on NFT marketplaces, first, we collect the addresses of the NFT marketplaces on Ethereum using Etherscan~\cite{etherscan_label}. Then, we study in which marketplaces the NFT transfer transactions occurred by looking at which smart contract address the transactions interact with. 

Tab.~\ref{tab:nftm_data} reports the six most important markets in the dataset. OpenSea is by far the largest, with 13.5\% of the NFTs and 30\% of the transactions in our dataset. LooksRare does not handle many transactions, but the volume is significant, almost 4B dollars, suggesting that they specialize in expensive NFTs. Many of the most valuable and important NFTs are traded in these marketplaces, but it is important to realize that trading (and wash trading) is also performed outside them.

\begin{table}
    \centering
    \small
    \caption{Data collected about NFTM}
    \label{tab:nftm_data}
    \begin{tabular}{l r r r}
    \toprule
        NFTM & NFTs  & Transactions & Volume(\$)\\ 
        \midrule
        OpenSea  & 4,686,022 & 6,979,112 & 14,292,698,925 \\
        LooksRare & 10,410 & 16,264 & 3,913,656,099\\
        Foundation & 132,675 & 212,973 & 140,952,188\\
        SuperRare & 20,384 & 27,637 & 59,441,772 \\
        Rarible & 19,886  & 22,171 &  45,982,785 \\
        Decentraland & 3,910 & 5,368 & 18,794,306\\
        \bottomrule
    \end{tabular}
\end{table}



\section{Methodology}

\subsection{Finding suspicious activities}
\label{subsec:graph_building}
The participants of a wash trading activity are a set of actors (in our case, accounts) that collude in buying and selling the same asset to inflate the asset's trade volume. Most often, the actors are actually the same entity. Thus, we are interested in finding suspicious transaction patterns that show a strong relationship between accounts trading the same NFT. As in previous works~\cite{chen2020traveling, von2022nft}, in a directed graph these suspicious transactions form closed cycles and strongly connected components (SCC).

Therefore, for each NFT~$i$ in our dataset, we build the \emph{transaction graph}, a directed multi-graph $\mathcal{G}_{i} = (V_{i}, E_{i})$. Set~$V_{i}$ has a node for every Ethereum account involved in at least one transaction of NFT~$i$. Edge $u\rightarrow v$ in $E_{i}$, $u,v\in V_i$, represents a transaction in which account~$u$ sells NFT~$i$ to account~$v$.
We annotate each edge in $E_{i}$ with the tuple $(t,h,s,p)$, where $t$ is the timestamp of the transaction, $h$ is the hash of the transaction, $s$ is the interacted smart contract, and $p$ is the amount paid for the NFT.

To start our analysis of suspicious activities, we compute all the strongly connected components in the transaction graphs consisting of at least two nodes and including single nodes with a self-loop. In the following, we will consider this peculiar definition of a strongly connected component. To find them, we use Tarjan's algorithm with Nuutila's modifications~\cite{nuutila1994finding} implemented by the Python library NetworkX~\cite{hagberg2008exploring}.
We found 905,562 NFTs with at least one strongly connected component with a total of 236,079 accounts. Out of these 905,562 NFTs, 874,079 have a strongly connected component of at least two nodes and 31,483 of only one node.

\subsection{Refining the graphs}
\label{subsec:cleaning_graph}

\textbf{Removing service accounts.}
Exchanges, CeFi services, and other cryptocurrencies services use EOAs to receive (send) funds from (to) their users. Moreover, some NFTMs offer escrow services using EOAs. 
The EOAs of these services interact with hundreds or thousands of user accounts, leading to the composition of strongly connected components that are not related to wash trading activities. Fortunately, Etherscan~\cites{etherscan_label}, the most popular Ethereum explorer, labels these accounts with the name of the service they belong to, allowing us to build a list of these specific accounts. We, therefore, collect all Ethereum addresses labeled as Exchanges, CeFi, and games by Etherscan.
Moreover, we add to the list the Ethereum Null Address (\texttt{0x0000}) since it can lead to the same problems as the EOAs described before. Indeed, it is common practice to use this address as the source of minting transactions (creation of an NFT) or as the recipient of a burn transaction (destruction of an NFT). 
We remove these nodes from the transaction graphs. After this process, the NFTs with a strongly connected component in the transaction graph reduced to 318,500, with a total of 128,505 accounts.


\textbf{Removing smart contract accounts.}
Strongly connected components could be accidentally created also by interaction with smart contracts. Indeed, as for service accounts, smart contract accounts often interact with many user accounts. Some examples of this kind of smart contract are those implementing DeFi services (\eg NFT-based games and liquidity pools using NFTs as LP tokens).
Thus, to avoid detecting false wash trading activities, we discard all smart contract accounts.
To ensure that we only exclude accounts associated with smart contracts and not those associated with externally owned accounts (EOAs), we only exclude accounts that contain bytecode. This is because smart contract accounts contain the bytecode that enables their functionality.
After this process, the NFTs with a strongly connected component are 305,314, with a total of 125,474 involved accounts.



\textbf{Removing zero-volume strongly connected components.}
The main goal of wash trading is to inflate the trading volume of an NFT. 
We define a zero-volume component as a component in which the NFT has been moved without any ERC-20 token or ETH transfer (\ie without payment).
In our set of graphs containing suspicious transactions, we find 292,158 strongly connected components where all the transactions have zero volume. While we keep zero volume transactions (they show a relationship between accounts), we discard these strongly connected components from our suspects. After this cleaning step, the NFTs with a strongly connected component are 13,156, involving 18,166 different accounts.

\subsection{Detecting wash trading}
To confirm that strongly connected components are actually wash trading, we look for patterns that show evidence of strong cooperation between the accounts.

\textbf{\ione{Zero-risk position.}}
According to the standard definition, wash trading is zero-risk market manipulation. Therefore, the idea is to suspect strongly connected components in the transaction graph that have zero balance after all the transactions and factoring out gas fees. Similar ideas have been used in~\cite{victor2021detecting,von2022nft}.
With this approach, we are able to confirm 3,121 strongly connected components as wash trading.

\begin{figure}
  \centering
  \includegraphics[width=.48\textwidth]{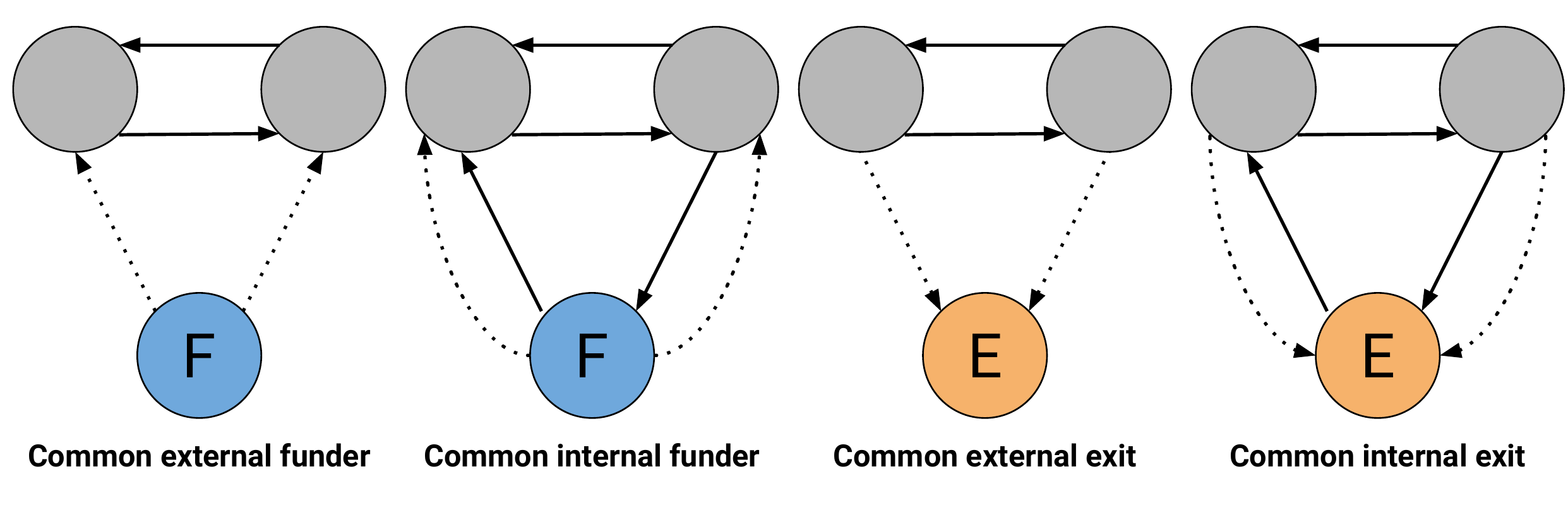}
  \caption{Examples of common funders and common exits. The dotted lines indicate transfers of ERC-20 tokens or Ether, while the solid ones represent an NFT transfer.} 
  \label{fig:funders_examples}
\end{figure}

\textbf{\itwo{Common funder.}}
\label{subsec:common_funder}
As done in previous
works~\cites{das2021understanding,xia2021trade} that focus on blockchain transactions analysis, clear evidence of collaboration between a set of accounts is the presence of an account that supplies funds to all or a subset of the alleged colluding accounts. 
We define a \textit{funding transaction} as a transaction that exclusively transfers Ethers or ERC-20 tokens to an account of the strongly connected component before the first transaction that moves the NFT in the colluding set of accounts. Moreover, we define the account that supplies the fund to the other accounts as \textit{common internal funder} if it belongs to the set of colluding accounts or as \textit{common external funder} if it does not.
The first and second images in~Fig.~\ref{fig:funders_examples} show an example of a common internal funder and a common external funder. The funder accounts are the blue nodes in the images. So, we detect wash trading if there is a common external funder that funds at least two different colluding accounts or if there is a common internal funder that funds at least one. In the case of a common external funder, we confirm the activity only if the account does not belong to the set of addresses of Exchanges and DeFi services collected in Sec~\ref{subsec:cleaning_graph}.
The wash trading activities detected with the common funder technique are 7,839, 1,579 external, and 6,260 internal.

\textbf{\ithree{Common exit.}}
Similarly to the previous approach, we can detect wash trading if the colluding nodes transfer their funds to a single account at the end of the operation. We define this account \emph{common external (internal) exit} as a node that receives funds from at least two accounts (one account) of the strongly connected component after the last transaction that moves the NFT.
The third and fourth images in Fig.~\ref{fig:funders_examples} show an example of a Common internal exit account and a Common external exit account. The exit accounts are the yellow nodes in the images. 
Also in this case, we discard the common external exit accounts of Exchanges or DeFi services.
The wash trading manipulations identified by the \emph{Common Exit} approach are 10,409, of which 7,384 with the Common inner exit account and 3,025 with the Common external exit account. 

\textbf{\ifour{Self-trade.}}
We consider the self-trade events as being verified de facto since it is the same entity that trades the NFT with itself, increasing the asset volume. We found 942 self-transfer events.

\textbf{\ifive{Leveraging confirmed wash trading events.}}
Finally, we use previously confirmed wash trading activities to validate additional wash trading events. 
In particular, we consider as wash trading a strongly connected component made of the same nodes of a previously strongly connected component identified as wash trading.
In this way, we can confirm 17 more wash trading activities, bringing the total number of wash trading events to 12,413.

\begin{figure}
  \centering
  \includegraphics[width=.45\textwidth]{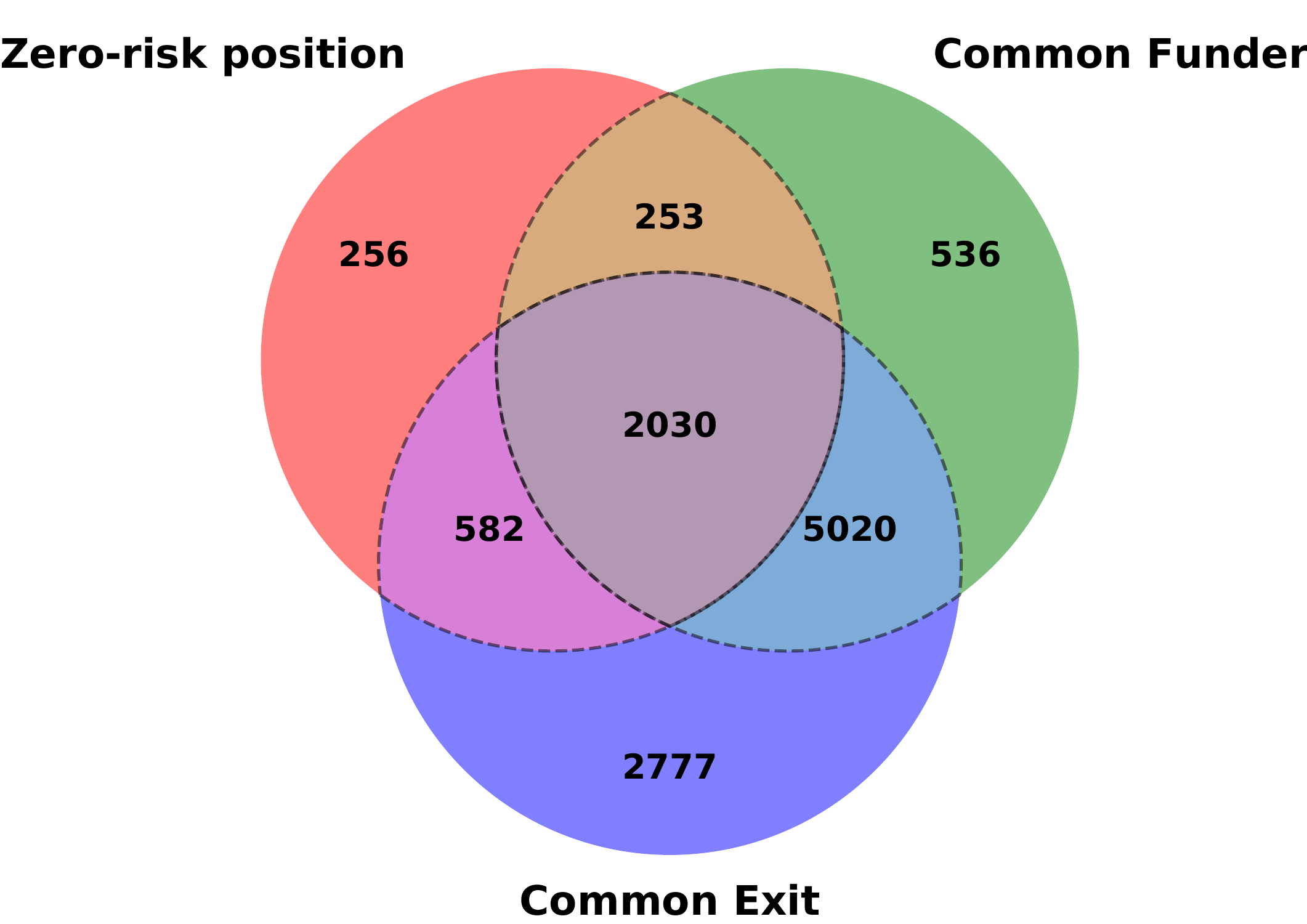}
  \caption{Venn diagram of the wash trading activities detected by each approach.} 
  \label{fig:methods_performance}
\end{figure}

\subsection{Comparison between methodologies.}
Using the zero-risk position, the common funder, and the common exit, we can confirm 11,454 wash trading activities. 
The Venn diagram in Fig.~\ref{fig:methods_performance} shows the number of wash trading activities detected by each approach. As we can see, more than half (56.79\%) of wash tradings are detected by both the common funder and common exit approaches. Therefore, a common pattern is as follows: The participants receive initial funds from a common account, perform the operation, and then the funds go back to a common account. 
Interestingly, 2,777 events are detected only using the common exit approach. We believe that, in this case, it is likely that the funder uses an exchange to provide the necessary money to the accounts. 
Indeed, we discovered 737 events where participants were funded by exchanges, with 375 of those activities coming from Coinbase~\cite{coinbase_site} and 276 from Binance~\cite{binance_site}. However, given the high number of transactions that exchange accounts perform, it is challenging to establish any relationship with high confidence.
Lastly, 2,030 (17\%) wash trading activities are detected by all three approaches based on transaction analysis, and more than 68\% of confirmed wash trading activities have been detected from at least two approaches.

\section{Results}

\subsection{Volume moved by wash trading activities}

\begin{table}
    \small
    \centering
    \caption{Data of wash trading on NFTMs}
    \label{tab:nftm_wash_trading}
    \begin{tabular}{l c r r}
    \toprule
        NFTM & \#NFT & Volume (\$) &  Total volume (\%)\\ 
        \midrule
        LooksRare & 534 & 3,318,450,936 & 84.79\%\\
        OpenSea  & 9,407 & 70,539,953 & 0.49\%\\
        Rarible & 189  & 1,206,082 & 2.62\%\\
        SuperRare & 30 & 147,166 & 0.24\%\\
        Decentraland & 20 & 33,028 & 0.17\%\\
        Foundation & - & - & -\\
        \bottomrule
    \end{tabular}
\end{table}

\begin{figure}
  \centering
  \includegraphics[width=.48\textwidth]{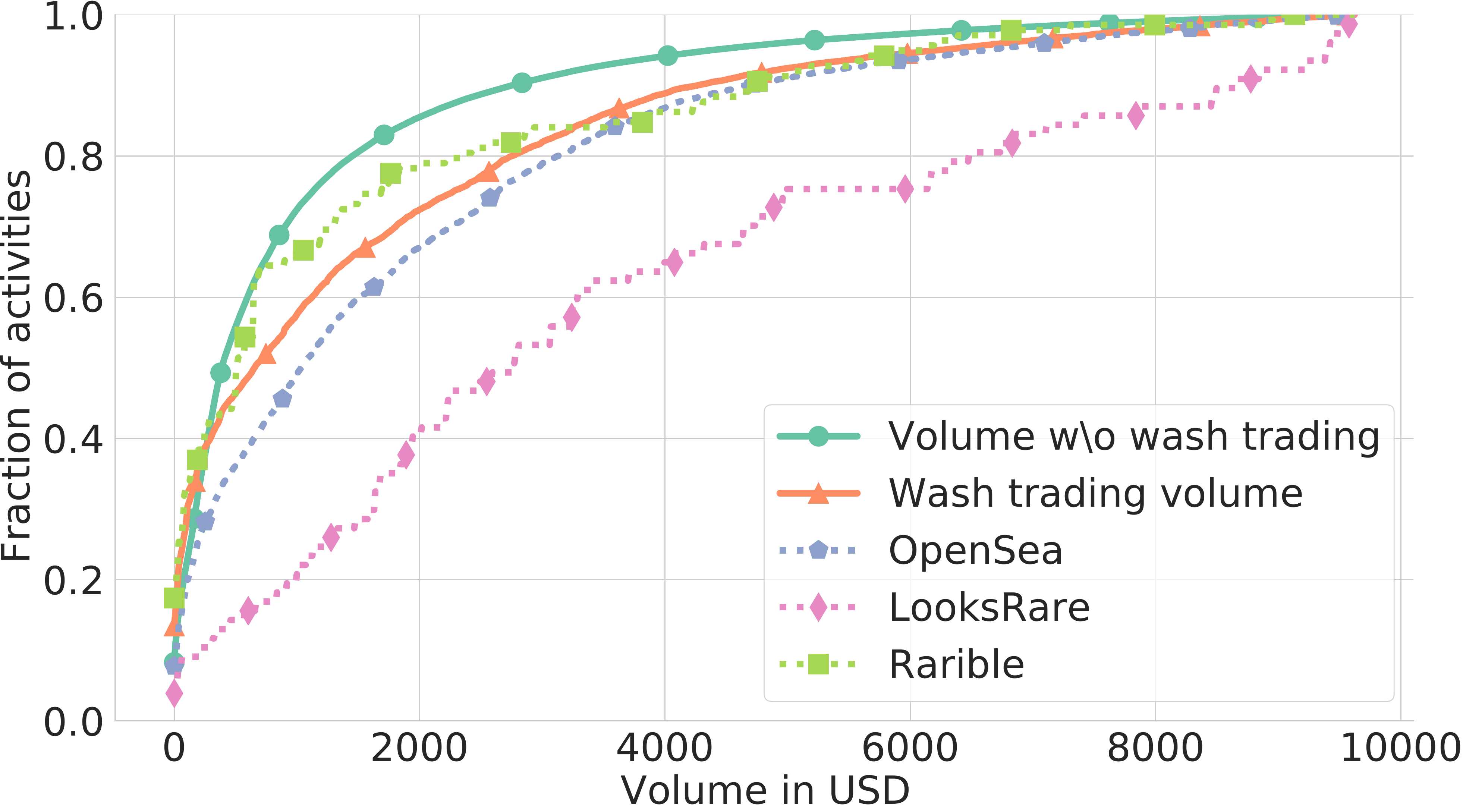}
  \caption{Wash trading volumes on different NFTMs.} 
  \label{fig:wash_trading_per_NFTM}
\end{figure}

Analyzing our data, we find 12,413 confirmed wash trading activities with a total volume of \$3,406,560,723. Tab.~\ref{tab:nftm_wash_trading} shows, for each NFTM, the number of NFTs affected by wash trading, the total wash trading volume on the marketplace, and the percentage of wash trading volume with respect to the total volume of the marketplace. As we can see, the marketplace most affected by wash trading is by far LooksRare, with a volume of \$3,318,450,936, which is 84.79\% of the total trading volume of the marketplace and 97.41\% of the whole wash trading volume in all the marketplaces.
Instead, Open Sea emerges as the marketplace with the largest number of wash trading operations, 9,407 out of 12,413 (75.78\%).
Nevertheless, the volume affected is just 0.49\% of its total transaction volume. 
Fig.~\ref{fig:wash_trading_per_NFTM} shows the CDF of the volume of wash trading activities in each marketplace and the volume generated by trading NFTs in the absence of market manipulation (solid green line). 
Looking at the CDF, it is evident that legit NFT trading generates much smaller volumes than NFT wash trading.
Moreover, the CDF highlights the higher volume generated in LooksRare compared to wash trading activities on other marketplaces. We investigate the reason for this anomaly in Section~\ref{sec:gains_analysis}.

Analyzing the volumes moved at the collection level, our investigation shows that the three collections more affected by wash trading are \textit{Meebits} with a volume of \$2,110,943,733 (78.22\% of its total volume), \textit{Terraforms} with \$496,049,600 (58.88\% of its volume), and \textit{Loot} with \$313,512,557 (49.75\% of its volume). Finally, we notice that all the wash trading activities on these collections have been carried out on the LooksRare marketplace.

\subsection{Temporal analysis}

\begin{figure}
  \centering
  \includegraphics[width=.48\textwidth]{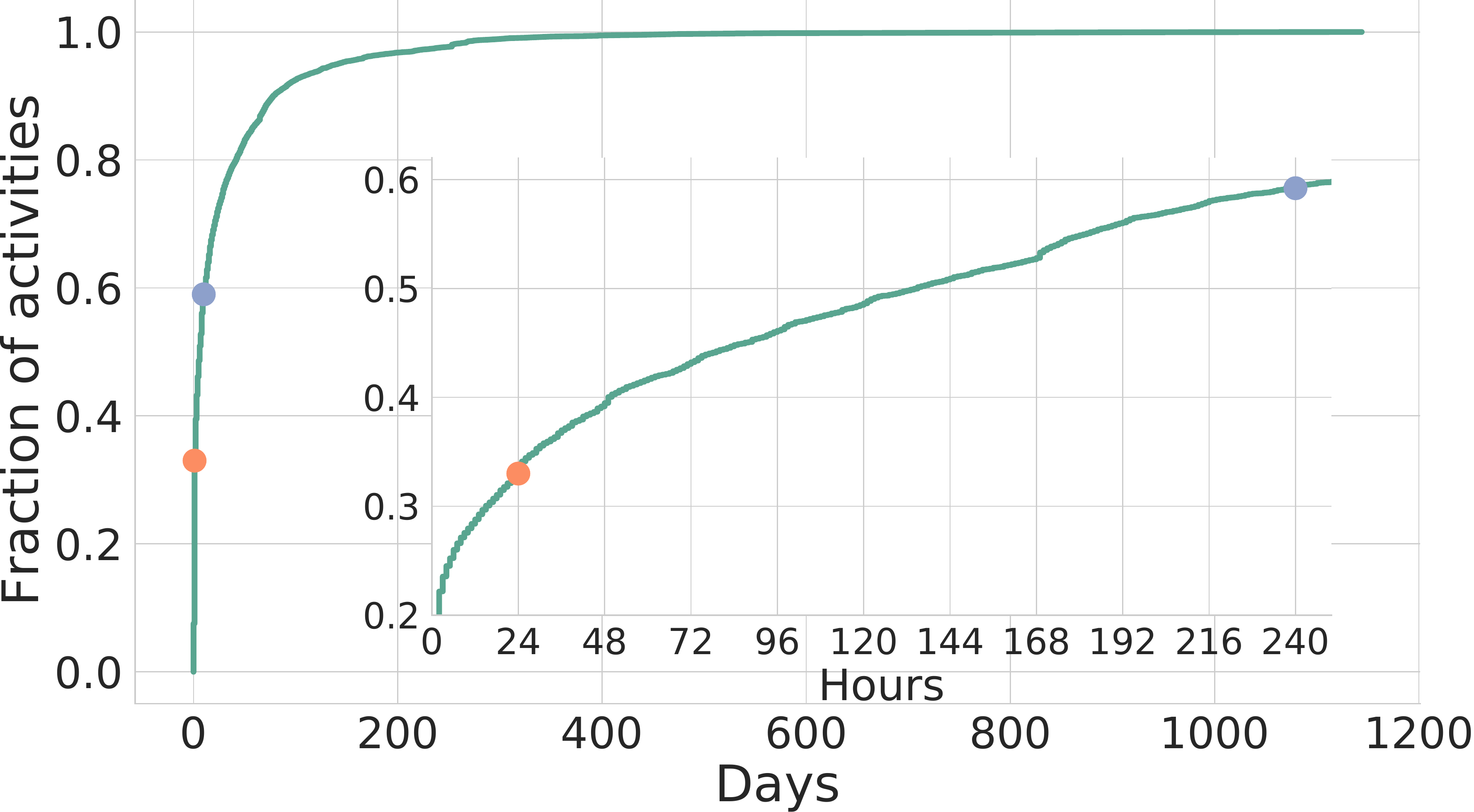}
  \caption{CDF of the lifetime of the wash trading activities. The orange dot shows the number of activities with a lifetime equal to or lower than one day, while the blue dot the number of activities with a lifetime equal to or lower than ten days.} 
  \label{fig:cdf_wash_trading_span}
\end{figure}

An intriguing aspect of wash trading is the time window of the typical activity. We define the lifetime of a wash trading activity as the time elapsed between the first and the last transactions between the accounts of the strongly connected component.
The CDF in Fig.~\ref{fig:cdf_wash_trading_span} shows the lifetime of wash trading activities. 
As we can see, approximately 33\% (4,157) of wash trading operations have a lifetime shorter than one day, the orange dot in the figure, while more than half of the activities (7,346) last less than ten days, the blue navy dot in the figure. Thus, in the majority of cases, wash trading appears to be a market manipulation that happens in a matter of days.

Then, we investigate the time that elapses between the wash trader buying the asset for the first time and the start of the wash trading activity. We see that 39\% (4,618) of the NFTs were purchased on the same day the manipulation started, and more than 75\% of them were purchased less than 14 days before. This result suggests that wash traders usually do not perform operations with NFTs they already own, but they buy the NFT only to perpetrate the fraud.

\begin{figure}
  \centering
  \includegraphics[width=.49\textwidth]{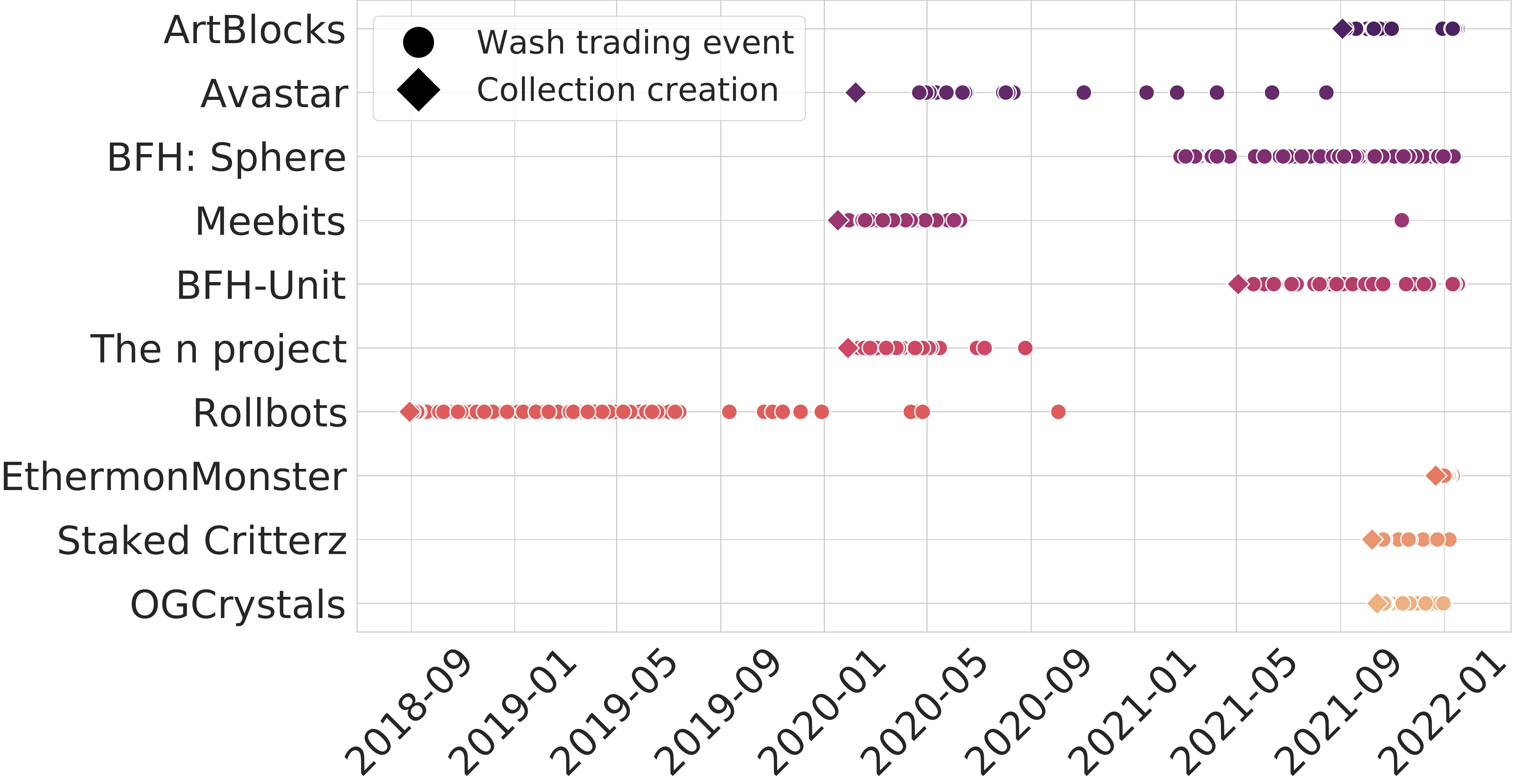}
  \caption{The circles indicate the occurrences of wash trading activities on the top 10 collections for the number of NFTs affected, while the diamonds represent the creation date of the collection.} 
  \label{fig:organized_wash_trading_events}
\end{figure}

\begin{figure}
  \centering
  \includegraphics[width=.48\textwidth]{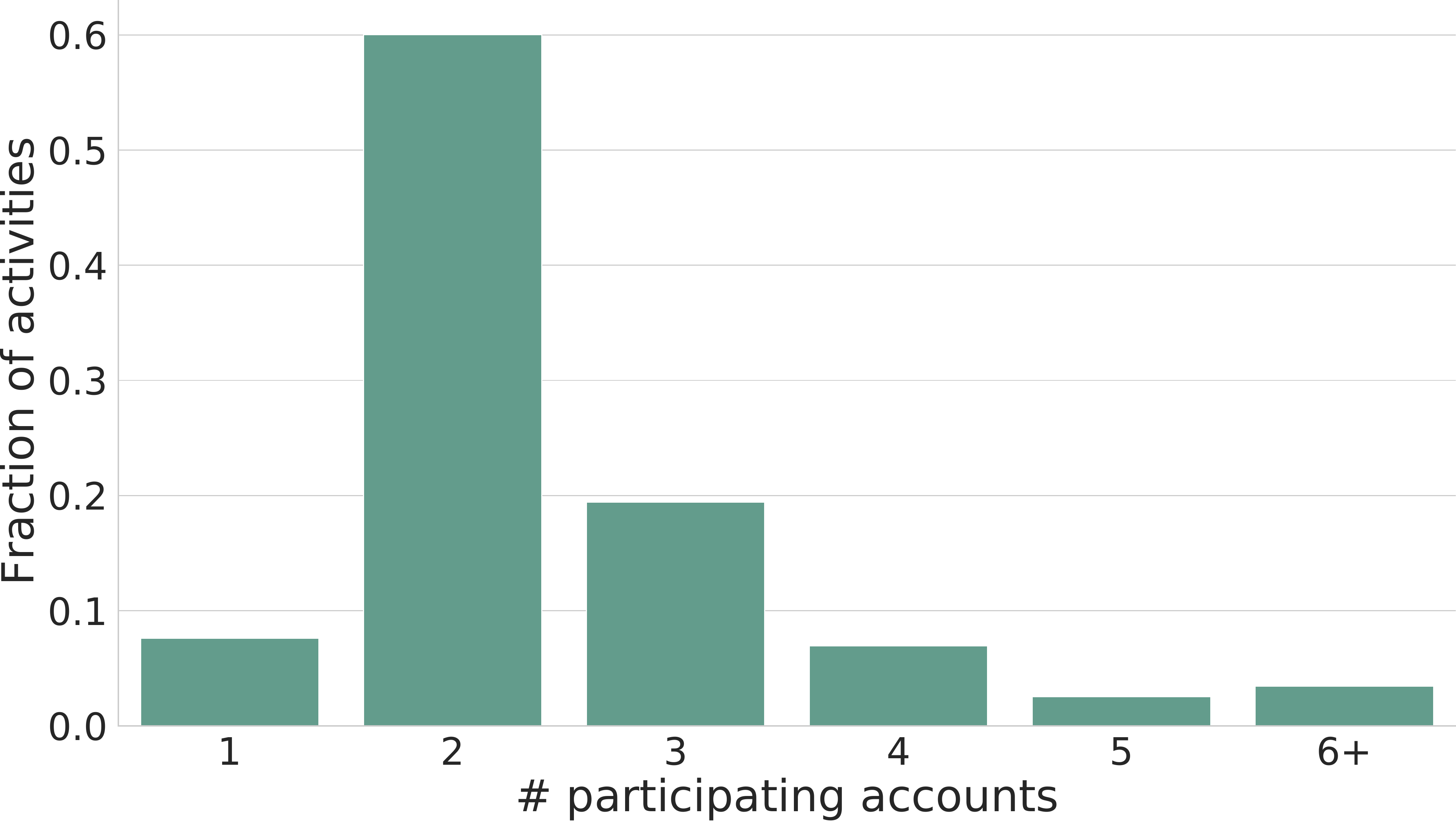}
  \caption{Number of accounts involved in wash trading activities.} 
  \label{fig:cdf_accounts_involved}
\end{figure}
Continuing on this aspect, Fig.~\ref{fig:organized_wash_trading_events} shows the occurrences of wash trading activities on the ten collections more affected by this phenomenon. As we can see, the majority of activities on NFTs occurred near the creation of the collection.

\begin{figure*}
  \centering
  \includegraphics[width=1\textwidth]{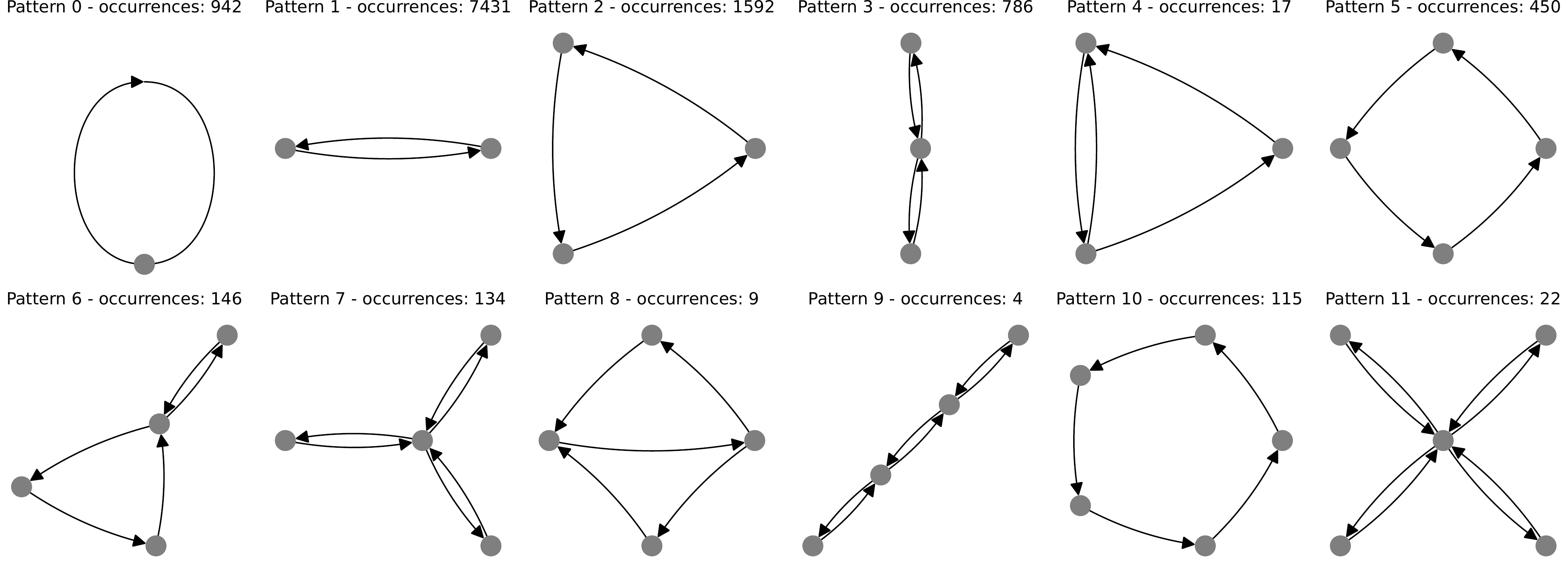}
  \caption{Most used patterns for wash trading activity. The patterns are ordered by the number of participants in the activities. These patterns represent more than 90\% of all the patterns we find.} 
  \label{fig:wash_trading_patterns}
\end{figure*}

\subsection{Patterns of wash trading}
\label{sub:patterns}
In this subsection, we analyze the number of accounts involved in wash trading activities and the pattern of transactions. 
Fig.~\ref{fig:cdf_accounts_involved} shows the distribution of the number of accounts in the confirmed wash trading activities, while Fig.~\ref{fig:wash_trading_patterns} the most common patterns of strongly connected components in the transaction graphs (covering 93.83\% of the wash trading activities) in our dataset.

As we can see, most wash trading activities (59.86\%) are performed by only two accounts doing ``round-trip'' trading of the NFT. 
Then, we find the strongly connected components with three accounts cover 19.42\% of our cases. Here, in the most common pattern with 1,592 occurrences, the wash traders circularly move the NFT. Looking at Fig.~\ref{fig:wash_trading_patterns}, we can notice that ``circular'' patterns (patterns 2, 5, and 10) are the most common, independently of the number of accounts involved in the activities---they are probably the most natural for wash traders.
Differently from wash trading on ERC-20~\cite{victor2021detecting} tokens, where almost all the manipulations are performed by a single account (96.15\%), in the case of wash trading on NFTs, they are only 7.6\%.
Arguably, the reason behind this difference lies 
in the restrictions imposed by some NFTMs, that do not allow the seller and buyer of a transaction to be the same.
Nonetheless, self-trade activities are relatively frequent. A reason could be that even if performed out of NFTMs, self-trades update the metric volume reported in NFT statistic services, as we notice investigating this kind of activity.

\subsection{Serial wash traders}
In this subsection, we investigate the presence of serial wash traders (\ie accounts participating in two or more wash trading activities). 
Of the 16,004 accounts involved in wash trading activities, we find that 4,348 (27.16\%) are serial wash traders, responsible for 9,054 (72.93\%) activities.
On average, a serial wash trader performs 4.85 activities, with the most active one involved in 655 operations.
This account 
mainly targets the Rollbots (RLB) collection\footnote{0x2f102E69cbcE4938CF7fB27ADb40fAd097A13668}, with 646 wash trades. 
In this case, the serial wash trader likely aims to increase the floor price (\ie the minimum price at which an NFT from the target collection can be purchased) of the Rollbots (RLB) collection. 
This behavior is common in most serial wash traders: 2,734 of them (62.87\%) perform multiple wash trading activities on NFTs of the same collection. 
Another interesting point about serial wash traders is their tendency to collaborate only with other serials, likely accounts belonging to the same entity. 
Indeed, 3,241 serial wash traders (74.54\% of them) participate in operations exclusively with other serials performing a total of 5,677 activities out of 9,054 (45.73\%).
The serial traders who performed the most activities together were two accounts that carried out 132 wash trading operations, all targeting NFTs belonging to the same collection, namely \textit{Avastar}\footnote{0xF3E778F839934fC819cFA1040AabaCeCBA01e049}.

\section{Gain or loss}
\label{sec:gains_analysis}
In this section, we analyze if wash trading is profitable and the typical gain the actors can expect from a wash trading operation.

There are two main ways to generate a profit with a wash trading manipulation. The first one is to exploit the token reward systems of some NFTMs, which reward a user with their token based on his trading volumes on the platform.
The second one is to artificially increase the value of an NFT and then try to sell it for a higher price.
In the following subsection, we analyze these behaviors separately.

\subsection{LooksRare and Rarible's reward systems}
LooksRare and Rarible promote their services by rewarding users of their marketplace with their tokens, the LOOKS, and the RARI tokens, respectively. 
These marketplaces distribute rewards to each user proportionally to his daily volume over the total daily volume of the marketplace. Hence, on each given day, the reward $R$ for a user $A$ is:

\begin{equation} \label{eq:looksrare}
    R_{A} = \frac{a}{b} \times c
\end{equation}
where, $a$ is the trading volume of $A$, $b$ is the total volume 
of the marketplace, and $c$ is the number of tokens available as a reward.
Of course, once they earn the rewards, users can swap the reward coins for other tokens (\eg{USDC, ETH, etc.}) using, for example, an exchange such as Uniswap. 

Clearly, this kind of reward system could be a gold mine for skilled wash traders, and, as we notice in Tab.~\ref{tab:nftm_wash_trading}, it was most probably largely exploited on LooksRare.
Thus, it is interesting to estimate the profits of wash trading in terms of rewards on these two platforms.
To estimate the profit, we compute the balance of a wash trading activity using the following formula:
\begin{equation} \label{eq:gain}
    balance = rewards - (NFTM_{fees} + Transaction_{fees})
\end{equation}
The formula subtracts from the rewards provided by the marketplace the costs to carry out the operation, the transaction fees ($Transaction_{fees}$) and the marketplace fees ($NFTM_{fees}$). 
To redeem the reward token from the NFTM, an account has to call the \textit{claim} function of particular smart contracts designed to distribute the reward token. Thus, we analyze all the transactions starting from one of the involved accounts and having as recipients the token distribution smart contract of LooksRare\footnote{0x453c1208B400fE47aCF275315F14E8F9F9fbC3cD} and Rarible\footnote{0x3b5d2B254224954547A33CbF753BcaA5eB4B27bd}. Then, for each account involved, we retrieve the number of tokens obtained from the first claim transaction that occurs after the activity.
Next, we compute the $rewards$ by summing the total tokens claimed by all participants' accounts.
Instead, to estimate the $Transaction_{fees}$, we sum all the gas fees paid by the involved accounts to perform the wash trading, including the fees to claim the reward tokens. Finally, for the $NFTM_{fees}$, we extract the amount of ETH transferred during the transactions to the treasury accounts (\ie accounts used to collect the platform's fees) of LooksRare\footnote{0x5924A28caAF1cc016617874a2f0C3710d881f3c1} and Rarible\footnote{0x1cf0dF2A5A20Cd61d68d4489eEBbf85b8d39e18a}. 
Since we deal with different tokens, we compute the balance by converting them into their corresponding value in USD on the day tokens were claimed or spent.
We consider the wash trading activity successful if the final balance is positive; otherwise, we consider it failed. 

Of the 534 confirmed wash trading activities on LooksRare, 75 do not claim the reward tokens. The same happened for 76 activities on Rarible.
A possible explanation for this behavior could be that the value of the fees to claim the reward tokens was higher than the value of the reward itself.
Excluding these activities from our statistics, in the end, we have 457 wash trading activities on LooksRare and 113 on Rarible.

\begin{table}
    \centering
    \caption{Token reward and wash trading}
    \label{tab:token_reward_wash_trading}
    \begin{tabular}{lrrrr}  
         & \multicolumn{2}{c}{LooksRare} & \multicolumn{2}{c}{Rarible}\\
        \cmidrule(r){2-3}  \cmidrule(r){4-5}
         &  Successful   &  Failed & Successful & Failed \\
        \midrule
        \# events & 365 & 92 & 107 & 6 \\
        min vol. (ETH) & 0.02 & 0.679 & 0.0002 & 0.019\\
        max vol. (ETH) & 70,823  & 44,970 & 28.70 & 30.75\\
        mean vol. (ETH) & 3,030 & 641 & 3.54 & 12.91\\
        max gain/loss (\$) & 32,437,888 & -213,913 & 289,217 & -4,269 \\
        mean gain/loss (\$) & 1,055,609 & -3,374 & 31,226 & -1,347 \\
        total gain/loss (\$) & 416,963,449 & -310,544 & 3,341,187 & -8,085\\
        \bottomrule
    \end{tabular}
\end{table}

Tab.~\ref{tab:token_reward_wash_trading} wraps up our results. The first thing that stands out is the high success rate. Indeed, almost 80\% of wash trading activities on LooksRare closed with a positive balance, and more than 93\% on Rarible.
On average, the trading volume on LooksRare (3,030 ETH) is much higher than the volume on Rarible (3.54 ETH), both in the case of successful and failed activities. Similarly, the mean gain and loss are more significant on LooksRare than on Rarible.
What is surprising is the total gain of successful operations, which is about \$416,652M and is far greater than the total loss of failed operations (\$310,544). Also, the mean gain of successful operations on LooksRare is exceptionally high (\$1.055M) compared to the average loss of unsuccessful activities (\$3,374). 


Moreover, we can see that the maximum volume among failed activities (44,970 ETH) on LooksRare is very close to the maximum volume among profitable ones (70,823 ETH) and significantly higher than the average (641 ETH).
Looking at Rarible, we find that the maximum volume among failed operations is even higher than the one among successful operations. 
Indeed, besides the volume generated, other reasons can cause the failure of the operations: The value of the reward tokens has a sudden drop, or the overall volume of the platform on the day of the activity was very high.

\subsection{NFTs sale}

The goal of wash trading is not only exploiting the reward systems of the marketplaces. The wash traders might try to drum up interest in the NFT to sell it at a higher price. This is particularly true on OpenSea, SuperRare, and Decentraland, which do not implement a reward system. Therefore, we now investigate the profits made by wash traders from this point of view.

We find that 7,564 out of 11,690 (64.7\%) activities in the three marketplaces are not followed by a transaction that sells the NFT to an external entity. In some cases, the NFT has been transferred for zero crypto, so these are probably just internal movements between accounts of the same owner.
It is possible that the wash traders attempted to sell some of these NFTs, listing them on NFTMs. However, we can not confirm this circumstance since the listing information is not written on chain. 
Furthermore, as we show in Sec.~\ref{sec:case_study} we discover cases in which wash traders exploit particular dynamics of some collections to enhance the NFT properties and rarity and, therefore, increase its value.

We find a sale transaction for 4,126 NFTs after the wash trading activity. In particular, 1,567 out of 4,126 (38\%) NFTs were sold on the same day the manipulation ended, whereas 75\% were sold within one month.
Simply comparing the buy price of the NFT by the wash trader with the re-selling price, we find that the large majority of the activities, around 65\% (2,682 out of 4,126), closed successfully, with a mean gain of 1.39 ETH and a maximum gain of 398 ETH. At the same time, the remaining 35\% (1,444) of events result in a loss with a mean of 0.59 ETH. The maximum loss we record is 45 ETH.

However, things change when we consider the fees. In this case, for each activity, in addition to the price to purchase and resell the NFT, we consider the transaction and NFTM fees spent by the participants of the wash trading activity.
Thus, the formula to compute the profit in this scenario is:

\begin{equation} \label{eq:resell_nft}
    balance = resell\_price - (buy\_price + fees)
\end{equation}

where $resell\_price$ is the payment of the resell of the NFT, $buy\_price$ is the price at which the wash trader purchased the NFT, and $fees$ includes the gas fees of the transactions and the NFTMs fees. 

Now, in 49.6\% (2,046) of the activities, the wash traders close the operation with losses, while in 50.4\% (2,080) with a profit. 
Here, the average loss is 0.65 ETH, while the average profit is 1.72 ETH. Including also the fees, it is not surprising that the average loss increases (from 0.59 ETH to 0.65 ETH). Instead, it is peculiar that the average profit increases (from 1.39 ETH to 1.72 ETH). 
This is because some wash trading activities that gained from the re-selling of the NFT failed to cover the gas fee due to the high number of transactions made.

Since the cryptocurrency market is very volatile, and the value of an asset can change dramatically in a few hours, we also analyze if the results change by computing the ETH value in USD at the time of the transactions. 
In this case, we find 1,875 (48.7\%) events for which the participants lose money with the wash trading activities and 1,972 (51.3\%) for which they profit.
This analysis suggests that wash trading operations are hazardous if the goal is to increase the asset price (success rate of around 50\%)---even if there is a chance to re-sell the NFT at a higher price, the transaction fee could be very expensive.

\section{Case study}
\label{sec:case_study}
In this section, we showcase some notable examples of the wash trading activities we found.

\mypara{\textbf{Token reward system exploit.}}
On LooksRare, two colluding accounts traded eight times the same NFT for an astonishing total volume of 5,733.884 ETH.
In particular, the operation lasts less than one hour, with only a few minutes between one sale and the next. 
Moreover, we notice that, at each trading iteration, the sale price decreases by the exact amount of fees of the previous transaction, causing a significant drop in the selling price (first sell at 930.314 ETH and last one at 690.314 ETH).
After the wash trading activity, both accounts claim the reward tokens for a total of 388,641.28 LOOKS (about 1,492,640.94\$ considering the token's value at the moment of claim). However, to compute the operation's balance, it is also necessary to consider its costs: The fee collected by LooksRare (114.65 ETH) and the gas fee paid for the transactions (0.356 ETH, of which 0.304 ETH for NFT sells and 0.052 ETH for claim transactions).
Thus, considering the price of ETH at the moment of the claim (about 3,373\$), the net gain is 1,104,722.25\$.

\mypara{\textbf{NFT resell.}}
On OpenSea, three accounts trade the NFT on a circular pattern (Pattern~2 in Fig.~\ref{fig:wash_trading_patterns}).
The volume of the operation is 19.8 ETH for a total value of more than \$65k. Throughout these trades, the price of the NFT increased from 0.66 ETH to 12.5 ETH. Finally, the wash trader sold the NFT for 14.85 ETH.
Given the low price at which the wash trader purchases the NFT (0.99 ETH), the total gain based on the difference between the purchase and selling price is 13.86 ETH (47,763 USD). However, even if we include the transactions and the marketplace (OpenSea) fees, the cost of this operation was just \$3.457, which brings the total profit of the operation to over \$44k with only three transactions, an investment return of more than 2000\%.
In contrast to the prior example, the operation lasted 64 days, with the initial transaction taking place on Jul 31 and the last on Oct 3. However, the final sale happened near the last trade of the operation, on Oct 6.
Contrary to the exploit of the reward systems, this kind of operation relies on external investors buying the NFT at a price proposed by the seller (\ie the one performing the wash trading operation).
For this reason, we believe more traditional wash trading on NFT is particularly risky.

\mypara{\textbf{Increasing rarity of the NFTs.}}
Lastly, we find some interesting cases in which an account trades multiple times, with a round-trip cycle (Pattern 1 in Fig.~\ref{fig:wash_trading_patterns}), different NFTs of the same collection. Interestingly, every time, the sales are performed by the \textit{same} account, while all the different buyers purchased the NFTs on OpenSea and then sent it back to the seller outside of the platform without any form of payment. Analyzing this scenario, we find that this kind of operation is performed because of how some NFT collections work. In this case, the targeted collection is \textit{OG:Crystals}.
The collection implements the following mechanism: When one of its NFTs is sold, it acquires additional and unique properties that might increase its value.
Therefore, this feature can incentivize a user to self-trade his NFT multiple times to increase his asset's value, generating an artificial interest in the NFT and the collection.


\section{Ethical considerations}
\label{sec:ethical_considerations}
In this work, we collect all ERC-721 NFT transactions of Ethereum to study the wash trading phenomenon.
All the data we analyzed are stored publicly on the blockchain, and EOA addresses are pseudo-anonymous. 
During our study, we never attempt to correlate users' addresses with external events jeopardizing the privacy of the users. 
Consequently, according to our IRB's policy, we did not need any explicit authorization to perform our experiments.

\section{Discussion}


\textbf{Can wash trading exist outside strongly connected components?}
In this work, we search for strongly connected components to detect possible candidates for the wash trading events. 
However, it is possible to perform wash trading without forming a strongly connected component, for example, moving the NFT always on different accounts. However, this trading pattern is very expensive for the wash trader and almost indistinguishable from standard trading, making any methodology easily prone to many false positives.



\textbf{Does wash trading require substantial capital?}
An NFT asset's volume can be significantly increased by exploiting \textit{Flash Loan}~\cite{wang2021towards}. 
Flash Loan is a particular kind of loan supported by several DeFi services (\eg AAVE). It is a particular loan that the user needs to repay in the same transaction in which it is taken.
A malicious user can produce a transaction that takes out the loan, transfers the NFT between two accounts he controls, and then repays the loan. Therefore, since wash trading is a zero-sum manipulation, capital is not a big concern.

\textbf{Why does Foundation not have wash trading?}
In our work, we did not find wash trading activities on the Foundation NFTM. We conjecture this is because of the high fee of the platform. Indeed, Foundation charges a total of 15\% of fees. Compared to OpenSea (2.5\%), LooksRare (2\%), and Rarible (2\%), they are noticeably higher. As a consequence, wash trading can easily generate more fees than benefits.

\textbf{Can marketplaces prevent wash trading activities?} As shown in this work, wash trading is strongly stimulated by the reward system adopted by some marketplaces.
We believe marketplaces can drastically reduce the phenomenon by denying rewards to accounts that take advantage of the wash trading. 
Moreover, by using the techniques described in this paper, marketplaces could flag suspicious transactions on the NFT page to warn users that wash traders might have artificially inflated the volume of the NFT.

\section{Related work}

\mypara{\emph{Data collection.}}
Collecting assets from the blockchains is still an open problem~\cite{cernera2022token}.
Different approaches have been proposed to collect Ethereum assets, in particular ERC-20. In~\cite{chen2020traveling, victor2019measuring}, they use the ABI interfaces to check for ERC-20 mandatory functions and events in the bytecode of smart contracts. Di Angelo et al.~\cite{di2021identification} use the same technique to identify ERC-20 as well as ERC-721 and others. These approaches are somewhat similar to ours, in which we identify transfer events that are ERC-721 compliant. Furthermore, it is crucial to notice that, for our study, we don't need to collect ERC-721 assets that have not been transferred between addresses. Indeed, since we want to measure wash trading events, assets without transfers cannot be subjected to this manipulation.

\mypara{\emph{Wash Trading.}}
The lack of regulation in the cryptocurrency ecosystem provides fertile ground for numerous typologies of market manipulations, such as pump and dump and front-running operations~\cites{gandal2018price, krafft2018experimental, la2020pump, daian2020flash}.
One of these manipulations is called wash trading. In the standard stock market, this has been extensively studied in different works~\cites{cao2014detecting,cao2015detecting,palshikar2008collusion,franke2008analysis}, but only recently this manipulation attracted attention in the crypto market.
Specifically, Victor et al.~\cite{victor2021detecting} analyze the wash trading behavior of ERC20 tokens in DEXs and propose a method to detect and quantify it based on data from IDEX and EtherDelta. They found that this manipulation accounted for a volume of more than \$159M.
However, in the crypto world, a particular type of asset gained attention in the last years: Non-Fungible Tokens (NFTs).
Das et al.~\cite{das2021understanding} give an overall assessment of the security issues in the NFT ecosystem. Among the issues, they also discussed NFTs wash trading. They collected the data from 7 marketplace APIs, web scraping, and Blockchain parsing from June 2021 to December 2021.
They built three graphs: A sales graph, a payment graph, and an asset transfer graph. They found the candidate searching for strongly connected components in the sales graph and confirmed the malicious activity if the accounts involved were connected inside the payment graph or the transfers graph. They found more than 9,000 instances of wash trading for a total volume of almost \$97M.
Recently, Von et al.~\cite{von2022nft} selected the top 52 NFT collections by volume from January 2018 to mid-November 2021 for one market, collecting the data from the blockchain.
They found alleged malicious activities for a total volume of \$149.5M.
Both these works analyze the phenomenon partially due to small-time windows or because they only target one market. In contrast, in our work, we analyze the phenomenon in the wild, collecting and studying all the assets ERC-721 transferred in Ethereum. 
More importantly, differently from the works mentioned above, we propose an in-depth characterization of the phenomenon, shedding light on wash trading's impact on the ecosystem and studying who performs the operations and why. Finally, we measure the manipulation profitability under different scenarios.

\section{Conclusion}

This work focuses on understanding the wash trading phenomenon of NFTs on the Ethereum blockchain. 
To the best of our knowledge, we are the first to explore wash trading activities on the whole blockchain, including transactions outside the NFTMs, and to 
characterize the following phase of this operation and its implication and profitability. 
We parse all transactions until January 2022, collecting 34,754,447 NFTs.
We applied and compared several techniques to detect wash trading activities, finding 12,413 operations.
Then, we thoroughly analyzed the activities we found from several perspectives.
We found that these activities usually last a few days, with 33\% of them lasting a single day and more than half less than ten days. Moreover, we discovered that the most common wash trading pattern, 59.86\% of activities, are performed by two accounts trading the NFT back and forth among themselves (round trip trading).
The wash trading activities we find account for a volume of more than \$3.4 billion. Nonetheless, we discovered that LooksRare generates 97.41\% of this volume, while wash trading scarcely affects other NFTMs. Investigating the reason for such intense wash trading activity on LooksRare, we find that it is related to their token reward system, which distributes platform tokens to users according to their trading volume.

Finally, we investigated whether wash trading activities on NFTs could be profitable by inflating the price due to manipulation or by exploiting the token reward systems. We learned that the resale of NFTs involves more risk (50\% of loss events), while the exploitation of the reward systems results in gain in more than 80\% of the events. 
As we saw, the transactions and NFTMs fees are one reason behind the failure of many wash trading activities. Thus, as future works, it is interesting to study the wash trading phenomenon on blockchains with low transaction fees, such as BNB Smart chain, Solana, or Polygon. 
Another possible direction is to use machine learning techniques, such as graph embedding, to detect wash trading patterns on the NFT graphs finding new patterns of interaction between nodes.


%


\ifCLASSOPTIONcaptionsoff
  \newpage
\fi



%
\printbibliography

%








\end{document}